\documentclass[reprint,amsmath,amssymb,aps,]{revtex4-1}
\usepackage{graphicx}
\usepackage{dcolumn}
\usepackage{bm}
\usepackage{color}
\usepackage{hyperref}
\usepackage{times} \usepackage{txfonts}
\usepackage[compact]{titlesec}
\titlespacing{\section}{0pt}{0pt}{0pt}
\hypersetup{colorlinks,citecolor=blue,urlcolor=blue}

\begin{document}
\title{Quantum spin Hall effect in strained (111)-oriented SnSe layers}
\author{S. Safaei}\thanks{safai@ifpan.edu.pl}
\author{M. Galicka}
\author{P. Kacman}
\author{R. Buczko}\thanks{buczko@ifpan.edu.pl}
\affiliation{Institute of Physics, Polish Academy of Sciences, Aleja Lotnik\'{o}w 32/46, 02-668 Warsaw, Poland}
\date{\today}

\begin{abstract}
Recently, the quantum spin Hall effect has been predicted in (111)-oriented thin films of SnSe and SnTe topological crystalline insulators \citep{NJP_Safaei,Liang_2015}. It was shown that in these films the energy gaps in the two-dimensional band spectrum depend in an oscillatory fashion on the layer thickness -- the calculated topological invariant indexes and edge state spin polarizations show that for films 20-40 monolayers thick a two-dimensional topological insulator phase appears. Edge states with Dirac cones with opposite spin polarization in their two branches are obtained for both materials. However, for all but the (111)-oriented SnTe films with an even number of monolayers an overlapping of bands in $\overline{\Gamma}$ and $\overline{\mathrm{M}}$ diminishes the final band gap and the edge states appear either against the background of the bands or within a very small energy gap. Here we show that this problem in SnSe films can be removed by applying an appropriate strain. This should enable observation of the Quantum Spin Hall effect also in SnSe layers.
\end{abstract}
\pacs{68.55.jd, 71.20.-b, 73.20.-r}

\maketitle

\section{Introduction}\vspace{4mm}
Quantum spin Hall (QSH) effect is a net result of oppositely polarized spin currents traveling in opposite directions along the edges of a two-dimensional topological insulator (2D TI). In 2D TI the metallic spin-polarized edge states are protected by time-reversal symmetry and robust against back scattering. Therefore the spin-polarized currents can flow without dissipation.
This is an extremely appealing feature for low-power-consumption usage, which triggered an ever growing effort to establish system that harbors QSH phase.

Lately, it has been theoretically predicted that for (111)-oriented thin films of SnTe and SnSe of particular thicknesses the QSH state can be obtained \cite{NJP_Safaei,Liang_2015}.
Both these materials are topological crystalline insulators (TCIs) whose surface states are protected not by time-reversal but by mirror plane symmetry \citep{Hsieh-NatCommun-2012,Dziawa-NatMater-2012}.
When the surfaces of the layer are well-separated, they exhibit four Dirac cones at the four projections of L points onto the 2D Brillouin zone of the (111) surface (one in $\overline{\Gamma}$ and three in $\overline{\mathrm{M}}$ points) \citep{PhysRevB.89.075317,PhysRevB_Safaei}. Otherwise, in ultrathin films, the surface states of top and bottom surfaces start to interact and energy gaps open where the Dirac nodes used to be.
These energy gaps can be modulated by changing the film thickness - it was shown that for some thicknesses the 2D valence and conduction bands intersect and the 2D states energy gap reaches negative values. In SnTe films only the one energy gap at $\overline{\Gamma}$ changes sign for some thicknesses, while in SnSe the three gaps at $\overline{\mathrm{M}}$ points oscillate with thickness. Thus, in both SnTe and SnSe layers the band inversions take place at odd number of points. In Ref. \cite{NJP_Safaei} it was shown that indeed a topological phase transition from trivial insulators to 2D TIs can be achieved by tuning the layer thickness. It is worth mentioning that a similar, thickness-induced phase transition  was obtained also for (001)-oriented ultrathin film, however, in this case the conversion is between trivial insulator and 2D TCI phase \citep{Ozawa,NatMat001_Fu}.

As it was discussed in Ref. \citep{NJP_Safaei}, for all but the (111)-oriented SnTe films with an even (ca 20) number of monolayers, the QSH effect would be difficult to observe because the $\overline{\Gamma}$ point and one of the $\overline{\mathrm{M}}$ points are projected to the same $k_x=0$ point in the 1D BZ of the [1$\bar{1}$0] edge. Due to the overlapping of these two projections the final band gap can be considerably diminished by mutual shadowing. As a result the edge states appear either against the background of the bands or within a very small energy gap.
In this article, we scrutinize the possibility of reducing such band gap shadowing in (111)-oriented SnSe films by biaxial strain. We predict that applying appropriate strain, e.g., by well chosen substrate, should enable observation of the QSH also in SnSe thin layers.

\vspace{5mm}\section{Band structure in strained layers}\vspace{4mm}
We investigate the effect of biaxial tensile strain on the energy spectrum, in particular on energy gaps, of (111)-oriented SnSe film. The film lays in the x-y plane and the z direction is chosen along [111] crystallographic axis. It is well known that strain changes the electronic properties of the material.  It can shift and split the degeneracy of energy levels. In heterostructures strain affects also the staggered band offsets and is considered as a powerful tool to engineer alternative routes for enhancing semiconductor structures \citep{Bauer-APL}.
For a homogeneous semiconductor, the variation of energy gap as function of the imposed strain is qualified in terms of deformation potentials tensor ($\mathcal{D}$) and strain component ($\varepsilon$).
When the deformation is not very strong and the distance between atoms is still close to the equilibrium lattice constant $a_0$, $\mathcal{D}$ can be regarded as a constant coefficient and hence, the change of the energy with strain is approximately linear.
Under isotropic biaxial strain in x-y plane, longitudinal $\varepsilon_{\parallel}$ and transverse $\varepsilon_{\bot}$ strain components are expressed as following:
\begin{align}
\varepsilon_{\parallel}&=\varepsilon_{x,y}=\frac{a_{x,y}}{a_0}=\frac{a_{\mathrm{D}}}{a_0}; &
\varepsilon_{\bot}=\varepsilon_{z}=\frac{a_{z}}{a_0},
\end{align}
where $a_{\mathrm{D}}=a_0(1+\mathrm{deformation}_{x,y})$, $a_z=a_0(1+\mathrm{deformation}_{z})$ and $a_0$ is the relaxed lattice constant. The longitudinal and transverse strain components are related through the Poisson's ratio $\nu$ that characterizes the elasticity of a crystal:
\begin{align}
\nu = -\frac{\varepsilon_{\bot}}{\varepsilon_{\parallel}}
\end{align}
The Poisson's ratio of rock-salt SnSe is assumed to be similar to the one for PbSe and taken from Ref. \citep{Zasavitskii} to be equal to 1.16 at low, $T\thickapprox 4\ K$, temperature. The lattice constant $a_0=5.97 \AA$ of the relaxed cubic SnSe was obtained in Ref.\cite{Wojek-PRB-2013} by first principle calculations.

\begin{figure}[h!]
\centering
\includegraphics[scale=0.8,width=0.5\textwidth,trim= 2.5cm 2cm 0 2cm]{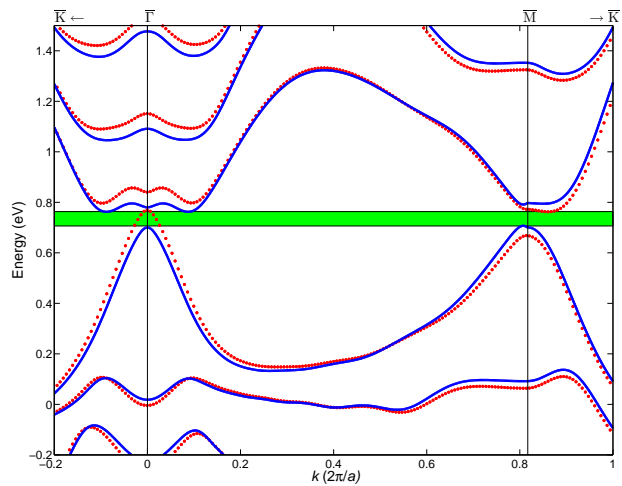}
\caption{Calculated band structure for relaxed (red dotted lines) and 1.14\% deformed (solid blue lines) anion-terminated 17-monolayers of SnSe.\label{Band}}
\end{figure}

\begin{figure}
\centering
\includegraphics[scale=0.7,width=0.5\textwidth,trim= 0.3cm 1.5cm 0.5cm 2cm]{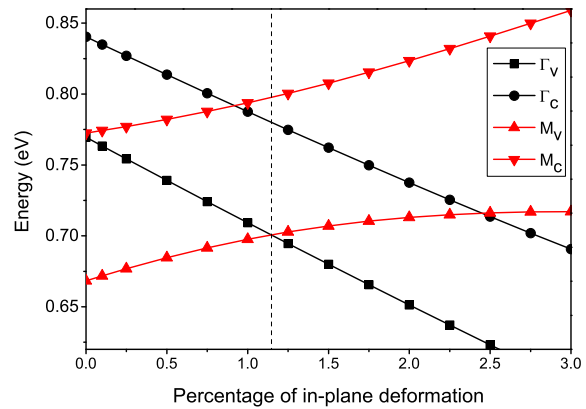}
\caption{The dependence of the energy gap of anion-terminated 17-monolayers of SnSe under biaxial strain on the percentage of deformation. The extrema of the valence band at $\overline{\Gamma}$ and $\overline{\mathrm{M}}$ are at the same energy level when the structure is $\sim$ 1.14\% deformed. $\Gamma_{\mathrm{v}}$, $\Gamma_{\mathrm{c}}$, $\mathrm{M}_{\mathrm{v}}$ and $\mathrm{M}_{\mathrm{c}}$ stand for valence and conduction band energies at $\overline{\Gamma}$ and $\overline{\mathrm{M}}$ points, respectively.\label{Egap_dfrm}}
\end{figure}

In Fig.~\ref{Band} we present the band structure of strained and fully relaxed 17-monolayer thick, anion-terminated (111)-oriented SnSe slab. The calculations were performed within a tight-binding approach. We have chosen this film to study the possibility of aligning the energy gaps by strain, since here the overlapping practically closes the energy gap. Indeed, as one can see in the figure, in the relaxed layer the energy gap at $\overline{\mathrm{M}}$ is completely covered by the valence band in the vicinity of $\overline{\Gamma}$ point. Our calculations show that while the deformation increases, the energy gap at $\overline{\Gamma}$ point is shifted down in energy, whereas the band gap at $\overline{\mathrm{M}}$ point is moving upward, as shown in Fig.~\ref{Egap_dfrm}. This is due to the difference between the effective masses in the vicinity of these two points of the Brillouin zone and subsequently different deformation potentials.

For deformation equal to $\sim$ 1.14\%, the tops of the valence band at $\overline{\Gamma}$ and $\overline{\mathrm{M}}$ have the same energy (compare Fig.~\ref{Egap_dfrm}) and the two energy gaps are aligned. As a result the band gap in the 2D spectrum of such slab opens -- the final total energy gap is equal to $\sim$ 0.57 meV, as shown in Fig.~\ref{Band} by the green strap.

The lattice constant $a_{\mathrm{D}}$ associated with the optimal deformation, which aligns the energy gaps in the anion-terminated 17-monolayers of SnSe, is equal to 6.04 \AA. Thus, the epitaxial growth of SnSe layers on a substrate whose lattice constant is equal to 6.04 \AA\ would lead to the desired deformation. The appropriate substrates that can provide such lattice mismatch are for example
PbS$_x$Se$_{1-x}$ with x$\simeq$0.46 and PbS$_x$Te$_{1-x}$ with x$\simeq$0.8.

\vspace{5mm}\section{Edge states and their spin polarization}\vspace{4mm}

\begin{figure}
\includegraphics[scale=0.4,width=0.5\textwidth]{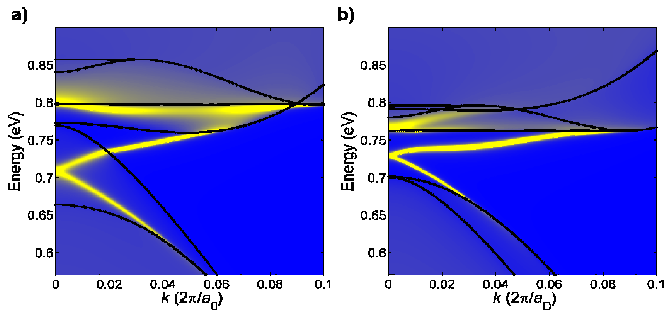}
\caption{The calculated spectral function of (a) relaxed and (b) 1.14\% deformed, 17-monolayers anion-terminated SnSe film at $k_x=0$. (The black solid lines denote the projection of the 2D bands boundaries.)\label{DOS}}
\end{figure}

 According to our previous results \citep{NJP_Safaei}, a 17-monolayers anion-terminated SnSe film should be a non-trivial insulator whose band gap inversion occurs at three $\overline{\mathrm{M}}$ points. Therefore, at the projections of the $\overline{\mathrm{M}}$ points on the [1$\bar{1}$0] edge, i.e., at $k_x=0$ and $k_x=\pm\sqrt{2}\pi/a$ points of the 1D BZ, three Dirac nodes are expected. However, the observation of these edge states would not be possible in the relaxed structures due to the mentioned above problem of strong shadowing of the band gap in such layer.
Here, to endorse our strain-induced alignment solution, we have calculated the [1$\bar{1}$0] edge spectral functions for both, relaxed and deformed, (111)-oriented anion-terminated SnSe films consisting of 17 monolayers.
Fig.~\ref{DOS}a, shows the edge states in the vicinity of $k_x=0$ for the relaxed structure. Notably the band gap is covered with the 2D valence band from $\overline{\Gamma}$ point. In turn, Fig.~\ref{DOS}b shows the edge state of the same film, which is now 1.14\% stretched. It evidently verifies that applying appropriate strain removes the shadowing of the gap. Now the edge state with a Dirac node can be clearly seen within the energy gap.

\begin{figure}
\includegraphics[scale=0.45,width=0.5\textwidth]{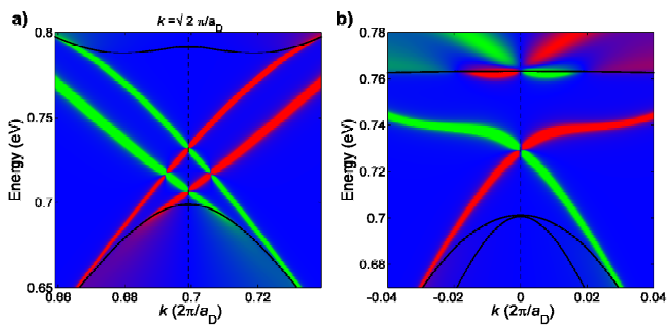}
\caption{Spin density of 1.14\% deformed 17-monolayers anion-terminated SnSe film in the vicinity of (a) $k_x=\sqrt{2}\pi/a_{\mathrm{D}}$ and (b) $k_x=0$. (Red and green lines denote the spin-down and spin-up polarization, respectively.)\label{Spin_DOS}}
\end{figure}

Ultimately, we have calculated the spin-up (G$\uparrow$) and spin-down (G$\downarrow$) contributions in the edge spectral function of the 17-monolayers anion-terminated SnSe film with 1.14\% deformation. The spin polarization of the edge states is defined as G$\uparrow$ $-$ G$\downarrow$.
The results are presented in Fig.~\ref{Spin_DOS}. As expected all three Dirac crossings with opposite spin-polarization in the two branches are now located inside the band gap. This should enable their observation.

\vspace{5mm}\section{Conclusion}\vspace{4mm}
In this article we show that biaxial strain can change the relative positions of various energy gaps in the 2D band spectrum of (111)-oriented SnSe ultrathin layers. For all such films, anion and cation terminated as well as films with even number of monolayers, a tensile strain appropriate to align the gaps at $\overline{\Gamma}$ and $\overline{\mathrm{M}}$ points has been found. For instance, 19 monolayers of SnSe, cation-terminated, requires 1.07 \% deformation to achieve an optimal alignment of the gaps, while 0.8 \% deformation is needed to cause the same in a 20-monolayer thick SnSe film. This means that the in-plane lattice constant of the film should be expanded to the value of about 6.02-6.04 \AA\ depending on the number of monolayers. The films can be thus stretched by using a substrate with such lattice constants. The substrate should be also a trivial insulator with
a wide band gap to ensure that the interface between the SnSe layer and the substrate simulates as much as possible a free surface. It seems that the best candidate for the substrate is PbS$_x$Te$_{1-x}$ with x$\simeq$0.8 or PbS$_x$Se$_{1-x}$ with x$\simeq$0.46 for films with an odd number of layers and  with x$\simeq$0.56 for 20-monolayer thick SnSe film. The energy gaps in the mentioned above compounds are ca 2-2.5 times bigger than in the film with odd number of monolayers, i.e., the band gap in the 17-monolayer thick SnSe
slab is equal to -0.1 eV while for PbS$_{0.46}$Se$_{0.54}$ $E_g$=0.21~eV and $E_g$=0.27~eV in PbS$_{0.8}$Te$_{0.2}$.
In the even number of SnSe monolayers case the difference between energy gaps of the film and substrate is larger -- for 20-monolayers the gap is equal ca -0.04 eV, what means that the energy gap 0.27 eV of PbS$_{0.56}$Te$_{0.46}$ is nearly 7 times bigger. It should be mentioned, however, that sulfur is usually not allowed in MBE systems and to grow such substrates other methods have to be applied. Finally, we note that the appropriate strain can be further adjusted by using the difference in thermal expansion coefficients of the SnSe and the used substrate. Of course one can think also about the possibility to obtain the needed deformation, i.e., the biaxial in-plane extension, by applying mechanically uniaxial pressure perpendicular to the (111) surface of the layer.
\vspace{5mm}\section*{Acknowledgements}\vspace{4mm}
This work is supported by the Polish National Science Center (NCN) Grants No. 2011/03/B/ST3/02659 and 2013/11/B/ST3/03934.

\bibliography{references}

\end{document}